\documentclass[conference]{IEEEtran}
\IEEEoverridecommandlockouts
\usepackage{cite}
\usepackage{tikz}
\usepackage{amsmath,amssymb,amsfonts}
\usepackage{algpseudocode}
\usepackage{algcompatible }
\usepackage{algorithm}
\usepackage{graphicx}
\usepackage{textcomp}
\usepackage{xcolor}
\usepackage{microtype}
\usepackage{float}
\def\BibTeX{{\rm B\kern-.05em{\sc i\kern-.025em b}\kern-.08em
    T\kern-.1667em\lower.7ex\hbox{E}\kern-.125emX}}
\begin{document}

\title{\textbf{SAC-AP}: \textbf{S}oft \textbf{A}ctor \textbf{C}ritic based Deep Reinforcement Learning for \textbf{A}lert \textbf{P}rioritization \\
%Deep Reinforcement Learning based Soft Actor-Critic method for Alert Prioritization

}
\author{\IEEEauthorblockN{Lalitha Chavali, Tanay Gupta, Paresh Saxena\\Dept. of CSIS, BITS Pilani\\ Hyderabad, India\\ \{p20190423, f20180343, psaxena\} @hyderabad.bits-pilani.ac.in  }

}

\maketitle

\begin{abstract}
Intrusion detection systems (IDS) generate a large number of false alerts which makes it difficult to inspect true positives. Hence, alert prioritization plays a crucial role in deciding which alerts to investigate from an enormous number of alerts that are generated by IDS. Recently, deep reinforcement learning (DRL) based deep deterministic policy gradient (DDPG) off-policy method has shown to achieve better results for alert prioritization as compared to other state-of-the-art methods. However, DDPG is prone to the problem of overfitting. Additionally, it also has a poor exploration capability and hence it is not suitable for problems with a stochastic environment. To address these limitations, we present a soft actor-critic based DRL algorithm for alert prioritization (SAC-AP), an off-policy method, based on the maximum entropy reinforcement learning framework that aims to maximize the expected reward while also maximizing the entropy. Further, the interaction between an adversary and a defender is modeled as a zero-sum game and a double oracle framework is utilized to obtain the approximate mixed strategy Nash equilibrium (MSNE). SAC-AP finds robust alert investigation policies and computes pure strategy best response against opponent's mixed strategy. We present the overall design of SAC-AP and evaluate its performance as compared to other state-of-the art alert prioritization methods. We consider defender's loss, i.e., the defender's inability to investigate the alerts that are triggered due to attacks, as the performance metric. Our results show that SAC-AP achieves up to 30\%  decrease in defender's loss as compared to the DDPG based alert prioritization method and hence provides better protection against intrusions. Moreover, the benefits are even higher when SAC-AP is compared to other traditional alert prioritization methods including Uniform, GAIN, RIO and Suricata.
\end{abstract}


\section{Introduction}
The need for cyber security is growing by the day to provide protection against cyber attacks \cite{biju2019cyber} including denial of service (DoS), distributed denial of service (DDoS), phishing, eavesdropping, cross site scripting (XSS), drive by download, man in the middle, password attack, SQL injection and malware. Further, intrusions due to unauthorized  access can compromise  the  confidentiality, integrity, and availability (CIA) triad of the cyber systems. To detect intrusions, intrusion detection systems (IDS) are used based on techniques like signatures, anomaly based methodologies and stateful protocol analysis \cite{liao2013intrusion}. 

%Recently, several  supervised  and  unsupervised  machine learning algorithms like support vector machines (SVMs), neural networks and random  forests have been applied for  intrusion  detection \cite{resende2018survey}, \cite{sinclair1999application}. However, these methods perform poorly on large  datasets  and  are incapable of detecting dynamic intrusions effectively due to their reliance on fixed characteristics of previous cyber attacks. To overcome the limitations of existing machine learning methods, several deep learning (DL) methods are integrated with IDS, for example, the use of recurrent neural networks (RNN) and deep neural networks (DNN) for intrusion detection \cite{chen2019ai}, \cite{roshan2018adaptive}. However, these methods lack adaptability to changing attack patterns. 

In this paper, we focus on alert prioritization \cite{alsubhi2008alert} using IDS. Specifically, IDS generates a high volume of security alerts to ensure that no malicious behaviour is missed. Several techniques exist for reducing the number of alerts without impairing the capacity of intrusion detection \cite{hubballi2014false}, \cite{salah2013model}. However, given the complexity of the environment and the numerous aspects that influence the generation of alerts, the quantity of alerts generated can be overwhelming for cyber security professionals to monitor, and hence it is necessary to prioritise these alerts for investigation. There exists various traditional approaches for prioritizing alerts like fuzzy logic based alert prioritization \cite{alsubhi2012fuzmet}, game theoretic prioritization methods like GAIN \cite{laszka2017game} and RIO \cite{yan2018get} but these methods show poor prioritization capabilities that lack estimating attacker's dynamic adaptation to the defender's alert inspection. 

Several  supervised  and  unsupervised  machine learning techniques have been also used for prioritizing alerts \cite{mcelwee2017deep}, \cite{sharafaldin2018toward}. However, these methods perform poorly on large datasets and are incapable of efficiently identifying dynamic intrusions due to their reliance on fixed characteristics of earlier cyber attacks and their inability to adapt to changing attack patterns. To address the challenges of a dynamic environment, reinforcement learning (RL) \cite{sutton2018reinforcement} has been used, where it learns through exploration and exploitation of an unknown and dynamic environment. The integration of DL with RL i.e., deep reinforcement learning (DRL) \cite{franccois2018introduction} has immense potential for defending against more sophisticated cyber threats and providing effective cyber security solutions \cite{nguyen2019deep}. Further, DRL based actor-critic methods \cite{sewak2019deep}, \cite{lazaric2007reinforcement}, \cite{saxena2020nancy}, \cite{naresh2022sac}, \cite{naresh2022deep} are considered as an efficient solution for dealing with continuous and large action spaces, implying its application in a dynamic cyber security environment \cite{lopez2020application}, \cite{shukla2021detection}.

Recently, \cite{tong2020finding} presents the framework that integrates adversarial reinforcement learning \cite{uther1997adversarial} and game theory for \cite{osborne1994course} prioritizing alerts. The interaction between an attacker (unauthorized user of a system) and the defender (who safeguards the system) is modeled as a zero-sum game, with the attacker selecting which attacks to perform and defender selecting which alerts to investigate. Furthermore, the double oracle framework \cite{mcmahan2003planning} is used to obtain the approximate mixed strategy Nash equlibrium (MSNE) of the game. Using this framework, \cite{tong2020finding} uses deep deterministic policy gradient (DDPG) \cite{lillicrap2015continuous} that computes pure strategy best response against opponent's mixed strategy. However, DDPG may entail instability due to its off policy nature and it may be prone to overfitting.

Specifically, most of the actor-critic methods like TRPO \cite{schulman2015trust}, PPO \cite{schulman2017proximal}, A3C \cite{mnih2016asynchronous}, etc., are on-policy RL methods that improve stability of the training but leads to poor sample efficiency and weak convergence. Furthermore, off-policy methods like DDPG \cite{lillicrap2015continuous} improves sample efficiency but it is sensitive to hyperparameter tuning and it is prone to the overfitting problem. In this paper, we present soft actor-critic based DRL algorithm for alert prioritization (SAC-AP), an off-policy method, that overcomes the limitations of DDPG. SAC-AP is based on a maximum entropy reinforcement learning framework that aims to maximize the expected reward while also maximizing the entropy.

The main contribution of the current work is to obtain robust alert investigation policies by using a soft actor critic for alert prioritization. We present the overall design of SAC-AP and evaluate its performance as compared to other state-of-the-art alert prioritization methods. Furthermore, we have presented the benefits of SAC-AP for two use-cases: fraud detection using fraud dataset \cite{link} and network intrusion detection using CICIDS2017 dataset \cite{sharafaldin2018toward}. We consider defender's loss., i.e., the defender's inability to investigate the alerts that are triggered due to attacks, as performance metric.  Our results show that SAC-
AP achieves up to 30\% decrease in defender’s loss as compared to
the DDPG based alert prioritization method \cite{tong2020finding} and hence provides
better protection against intrusions. Moreover, the benefits are
even higher when SAC-AP is compared to other traditional
alert prioritization methods including Uniform, GAIN \cite{laszka2017game}, RIO \cite{yan2018get} and
Suricata \cite{link2}.

The rest of this paper is organized as follows. A brief background of DRL based actor-critic methods in the context of our work is described in Section \uppercase\expandafter{\romannumeral 2 \relax}. The system architecture is described in Section \uppercase\expandafter{\romannumeral 3 \relax}. Our  proposed approach, SAC-AP to find the approximate best responses are presented in Section \uppercase\expandafter{\romannumeral 4 \relax}. The experiment results are provided in Section \uppercase\expandafter{\romannumeral 5 \relax}. Section \uppercase\expandafter{\romannumeral 6 \relax} concludes the paper.

%\section{Related Work}

\section{Background}
In RL, an agent interacts with the environment and at each time step $t$, it observes  state $s_t \in S$ and takes an action $a_t \in A$.  The action selection is based on the chosen policy $\pi(a|s)$ that is a mapping from state $s$, to the action $a$ to be taken in a given state. The agent then receives a reward $R_{t+1}$ and moves to next state $s_{t+1}$ based on action $a_t$ \cite{sutton2018reinforcement}. The goal in RL is to maximize the total discounted reward and to obtain optimal policy. Let us also denote the value function, $V_\pi(s)$, as a function of discounted reward, specifying  how good to be in a particular state $s$. It is given by, 
\begin{equation}\label{eq2}
    V_\pi(s) = E_\pi\Bigg[\sum_{k=0}^{\infty}\gamma^kR_{t+k+1}|s_t=s\Bigg]
\end{equation}
where $V_\pi(s)$ is value function for a state $s$ and policy $\pi$, and $\gamma\in[0,1]$ is the discount factor.

Similarly, the action value function $Q_\pi(s,a)$ is defined as the value of  taking action $a$ in a state $s$ under policy $\pi$:
\begin{equation}\label{eq2}
    Q_\pi(s,a) = E_\pi\Bigg[\sum_{k=0}^{\infty}\gamma^kR_{t+k+1}|s_t=s, a_t=a \Bigg]
\end{equation}

The optimal $Q(s,a)$ is calculated as follows:
\begin{equation}\label{eq2}
    Q^*(s,a) = \max_{\pi\in\Pi} Q_\pi(s,a) 
\end{equation}

We can say that the policy $\pi$ is better than the policy $\pi^{'}$, if $Q_{\pi}(s,a)$ is greater than $Q_{\pi^{'}}(s,a)$ for all the states and action pairs. The optimal policy $\pi^{*}$ is given by,

\begin{equation}\label{eq2}
    \pi^{*}(s) =  \textit{argmax}_{a\in A}( Q^{*}(s,a)) 
\end{equation}

In reinforcement learning, value based methods obtain optimal policy by choosing the best action at each state using the estimates generated by an optimal action value function. Some of the value based methods include SARSA, Q-learning, and DQN \cite{sutton2018reinforcement}. These methods  are not suitable for large and continuous action spaces like robotics. Furthermore, the policy based methods are another type of reinforcement learning methods that directly optimize a policy based on policy gradient like REINFORCE algorithm \cite{sutton2018reinforcement} without using any explicit value function. These methods work well for continuous action space but are prone to high  variance and slow convergence. To overcome these limitations of value based and policy based methods, actor-critic method \cite{sewak2019deep} are introduced that combine value and policy based methods into a single algorithm.

\subsection{Overview of actor-critic methods}

\begin{figure}[t]
   \begin{center}
    \includegraphics[scale=0.5]{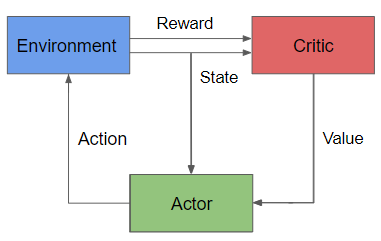}
    \end{center}
    \caption{Actor Critic Framework}
    \label{picture1}
\end{figure}

Actor-Critic (AC) methods utilize two different models: actor and critic as shown in Figure \ref{picture1}. The actor model takes the current environment state as input and gives the action to take as output  \cite{sewak2019deep}.  The critic model is a value based model that takes state and reward as inputs and measures how good the action taken for the state using state value estimates. Actor-Critic methods combines the advantages of both value and policy based methods that include learning stochastic policies, faster convergence and lower variance. The advantage function specifies how good it is to be in that state. If the rewards are better than the expected value of the state, the probability of taking that action is increased. The advantage function for state $s$ at time $t$ is calculated as follows: 
\begin{equation}\label{eq1}
    \delta_t = R_t + \gamma V(s_{t+1};W_{t}) - V(s_t;W_t)
\end{equation}
where $R_t$ is the reward at time $t$,  $\gamma$ is the discount factor, $V(s_{t+1};W_{t})$ is the state value estimator function parametrized by weight vector $W$, which is updated as follows:
\begin{equation}\label{eq}
    W_{t+1} = W_t + \alpha_w\delta_t\nabla_wV(s;W_t)
\end{equation}
with $\alpha_w$ as the learning rate. Similar to the critic model, the policy estimator function of an actor model with policy $\pi$ is also updated in each iteration by updating it's parameter vector $\theta$ using the following equation:
\begin{equation}
    \theta_{t+1} = \theta_t + \alpha_\theta \delta_t\nabla_\theta log\pi_\theta(a_t|s_t;\theta_t)
\end{equation}
The actor-critic method converges and obtains the optimal policy by following the updates described in this section.

\subsection{Overview of Soft Actor-Critic (SAC)}
Soft actor critic method is an off policy, actor-critic deep
reinforcement learning algorithm. SAC is based on maximum entropy reinforcement learning framework where actor aims to maximize the expected reward while also maximizing the entropy. SAC is stable, highly sample efficient, provides better robustness and solves the problem of overfitting and hence performs better than other state-of-the-art actor-critic methods including DDPG. SAC aims to optimize the maximum entropy objective function as follows \cite{haarnoja2018soft}: 
\begin{equation}
   \pi^{*}(.|s_t) = argmax _\pi E_\pi\Bigg[\sum_{t} R_t + \beta H\Big(\pi(.|s_t)\Big)\Bigg]
\end{equation}
where $H(\pi(.|s_t))$ is the entropy of the policy for the visited state under that policy $\pi$. The entropy is multiplied by the coefficient $\beta$ where if $\beta$ value is zero then the optimal policy is deterministic or else the optimal policy is stochastic.  The soft policy iteration includes evaluating the policy using the following Bellman equation that converges to the optimal policy with $\tau^\pi$ as the bellman backup operator, where $\tau^\pi$ is given by,
\begin{equation}
    \tau^{\pi}Q(s_t,a_t) \displaystyle\leftarrow R_t + E_{s_{t+1}\sim p}\Big[V(s_{t+1})\Big]
\end{equation}
where $p$ is the probability density of $s_{t+1}$ given action $a_t$ and current state $s_t$ induced by policy $\pi$.
The value function is calculated using the following soft bellman iteration:
\begin{equation}
    V_{s_{t}} = E_{a_{t}\sim \pi}\Bigg[Q(s_{t},a_{t}) - \beta log \pi(a_{t}|s_{t})\Bigg]
\end{equation}
The policy is updated towards the exponential new soft Q function using soft policy improvement step as follows:
\begin{equation}
    \pi_{new} = argmin_{\pi^{'}\in\Pi} D_{KL}\left (\pi^{'}(.|s_t)||\frac{exp \Big(\frac{1}{\beta} Q^{\pi_{old}}(s_t,.)\Big)}{Z^{\pi_{old}}(s_t)} \right )
\end{equation}
where $Z^{\pi_{old}}(s_t)$ normalizes the distribution and we have $Q^{\pi^{new}}\geq Q^{\pi^{old}}$ for the new policy that leads to the sequences of policies with the Q values increasing monotonically. In addition to using function approximators for both Q-function and the policy, SAC uses a state value function to approximate the soft value to stabilize training.

\section{System Architecture}
\begin{figure*}[t]
   \begin{center}
    \includegraphics[scale=0.5]{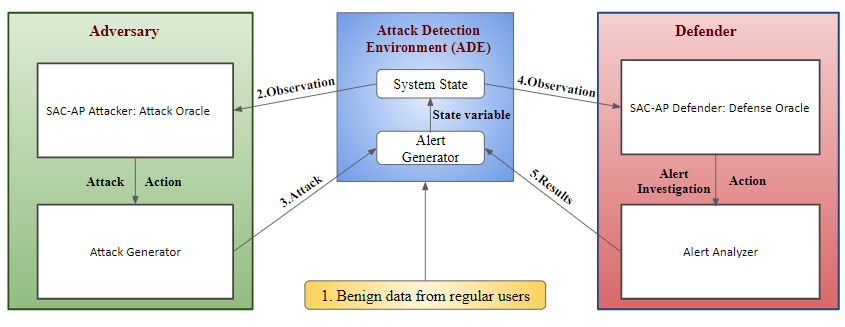}
    \end{center}
    \caption{System Architecture}
    \label{picturee}
\end{figure*}

The system architecture is divided into four modules: regular users, attack detection environment (ADE), adversary, and defender, as shown in Figure \ref{picturee}. The regular users are the system's registered users. The adversary is the system's unauthorised user who intends to launch various forms of attacks. ADE generates alerts as a result of an adversary's actions, but on numerous occasions, alerts are also generated as a result of normal user activity. The defender's role is to safeguard the system by investigating the alerts generated by the ADE.

ADE contains system state as presented in Table \ref{table1}. It includes $N_t^k$, i.e., the number of alerts of type $t$ that have not been investigated by the defender at time $k$. We assume that both the defender and the attacker are aware of the value of $N_t^k$. Additionally, the system state contains $M_a^k$, a binary value that indicates whether an attack $a$ was executed, and it is only known by the attacker. Furthermore, the system state also includes $S_{a,t}^k$, which denotes the number of alerts of type $t$ that were raised due to an attack $a$, and it is also known only by the attacker. %Alerts are raised for the observed suspicious behaviour by the alert generator. The system generates alert counts for each alert type at the end of each time period. According to a probability distribution $P$, stochastically alerts are triggered for each attack action. Benign alerts are generated according to the known distribution $F$.  

Following that, the adversary's knowledge include system's state at time period $k$ and defender's alert investigation policies in the previous rounds. The number of attacks the adversary can execute is limited by it's budget constraint $D$, and the adversary must adhere to the following constraint:
\begin{equation}
\sum_{a\in A} \alpha_{-1,a}E_a\leq D
\end{equation}
where $\alpha$ is a player's action, $D$ is an adversary's budget and $E_a$ is a cost of mounting attack $a$. The adversary is represented with two modules, one of which is an attack oracle and the other is an attack generator. The attack oracle runs a policy at each time period $k$ that maps state to choice of subset of attacks to be executed by the attack generator. The adversary succeeds if the defender does not investigate any of the alerts generated as a result of an attack.

Additionally, the number of alerts the defender can choose to investigate is limited by it's budget constraint $B$, where the defender should follow the following constraint:
\begin{equation}
    \sum_{t\in T} \alpha_{+1,t}^{k}C_t\leq B
\end{equation}
where, $\alpha$ is a player's action, $B$ is a defender's budget and $C_t$ is a cost of investigating alert type $t$. The defender consists of two modules: defense oracle and alert analyzer. At time $k$, the defense oracle runs a policy, that maps state to choice of subset of alerts to be investigated by the alert analyzer. 

%In the following section we describe our proposed SAC-AP DRL approach, that models the game theoretic interaction between an attacker and the defender that aims in computing robust alert prioritization policy.

\begin{table}[hbtp]\centering
 \caption{System state at time period $k$}
 \label{table1}
 \begin{tabular}{ |p{1.5cm}|p{2.5cm}| }
 \hline
 Symbol & Description \\
 \hline
 $N_t^k$ & Number of alerts of type $t$, not being investigated by the defender (Observed by both defender and attacker)\\
 \hline
 $M_a^k$ & Binary indicator that represents whether attack $a$ was executed (Observed only by attacker)\\
 \hline
$S_{a,t}^k$ & Number of alerts of type $t$ raised due to attack $a$ (Observed only by attacker) \\
\hline
\end{tabular}
\end{table}

\section{Proposed Solution: SAC-AP}

In this section, we present the overview of the solution and the proposed soft actor critic based alert prioritization (SAC-AP) algorithm.

\subsection{Solution Overview}
The interaction between an adversary and the defender (denoted by $-1, +1$) is modeled as a two player zero-sum game. In a given state, the attacker's set of possible actions includes all subsets of attacks and defender's set of possible actions includes all subsets of alerts satisfying the budget constraints mentioned in the previous section. The attacker's policy maps state to the subset of attacks and the defender's policy maps state to the subset of alerts.  We consider player's deterministic policies as pure strategies and stochastic policies as  mixed strategies. Let $\Pi_v$ and $\sum_v$ denotes the set of player's pure strategies and mixed strategies, respectively. Let us also denote $U_v(\pi_v,\pi_{-v})$ as the utility of player $v$ where $\sum_{v\in(-1,+1)}U_v(\pi_v,\pi_{-v}) = 0$ since its a zero-sum game. 

%As this is a two player zero sum game the summation of utilities of an attacker and the defender corresponds to zero i.e., one player's win will be  equal to the other player's loss. For example, consider two players choosing pure strategies to play, then $\sum_{v\in(-1,+1)}U_v(\pi_v,\pi_{-v})$ = 0 where $U_v(\pi_v,\pi_{-v}$ denotes utility of player $v$.The equations for calculating utilities of the players for their corresponding pure and mixed strategies are taken form the work \cite{b6}.

The expected utility when a player $v$ selects pure strategy $\pi_v\in\Pi_v$ and it's opponent selects mixed strategy $\sigma_{-v}\in\sum_{-v}$ is given as follows \cite{tong2020finding}:

\begin{equation}
    U_v(\pi_v,\sigma_{-v}) = \sum_{\pi_{-v}\in\Pi_{-v}} \sigma_{-v}(\pi_{-v})U_v(\pi_v,\pi_{-v})
\end{equation}
Further, player $v's$ expected utility when it selects mixed strategy $\sigma_v\in\sum_v$ and it's opponent also selects mixed strategy $\sigma_{-v}\in\sum_{-v}$ is given by:
\begin{equation}\label{eqq}
    U_v(\sigma_v,\sigma_{-v}) = \sum_{\pi_v\in\Pi_v} \sigma_{v}(\pi_{v})U_v(\pi_v,\sigma_{-v})
\end{equation}
Using the above equations, the utility of the defender when it chooses pure strategy $\pi_{+1}\in\Pi_{+1}$ and it's opponent (attacker) also chooses pure strategy $\pi_{-1}\in\Pi_{-1}$ with discount factor $\gamma$, is given by:
\begin{equation}
 U_{+1}(\pi_{+1},\pi_{-1}) = E\bigg[\sum_{k=0}^{\infty}\gamma^k  R_{+1}^{(k)}\bigg]
\end{equation}

where, $R_{+1}^{(k)}$ is the defender's reward at time period $k$, and it is given by: 
\begin{equation}
    R_{+1}^{(k)} = -\sum_{a\in A}L_a  \hat M_a^{(k)}
\end{equation}

with $L_a$ as defender's loss when attack $a$ is not detected and $\hat M_a^{(k)}$ is a binary indicator when the executed attack has not been investigated. The utility of the attacker is given as, $U_{-1}(\pi_{+1},\pi_{-1}) = -U_{+1}(\pi_{+1},\pi_{-1})$.

To solve this zero-sum game, we use an extension of double oracle algorithm \cite{tsai2012security}.  The goal is to compute MSNE to find robust alert investigation policy. The MSNE  condition for two players with mixed strategy profile $(\sigma_v^*,\sigma_{-v}^*)$ is given as follows:
\begin{equation}
    U_v((\sigma_v^*,\sigma_{-v}^*)) \geq U_v((\sigma_v,\sigma_{-v}^*)) \forall \sigma_v\in\Sigma_v
\end{equation}

where the MSNE of a game is computed by solving the following linear program:
\begin{align*}
    \textit{ max } U_v^*\\
    \textit{ s.t. } U_v(\sigma_v,\pi_{-v}) \geq U_{v}^{*}, \forall \pi_{-v}\in\Pi_{-v}\\
    \sum_{\pi_v\in\Pi_{v}}\sigma_v(\pi_v) = 1\\
    \sigma_v(\pi_v)\geq0, \forall\pi_v\in\Pi_v
\end{align*}
where the mixed strategy provisional equilibrium is obtained by solving the above mentioned linear program and by considering the set of player's random policies $(\Pi_{+1},\Pi_{-1})$. The attack oracle then computes adversary's best response $\pi_{-1}(\sigma_{+1})$ to the equilibrium mixed strategy $\sigma_{+1}$ of the defender. Similarly, the defense oracle computes defender's best response $\pi_{+1}(\sigma_{-1})$ to the equilibrium mixed strategy $\sigma_{-1}$ of the attacker. Furthermore, these best responses are then added to $(\Pi_{+1},\Pi_{-1})$ policy sets. The process repeats until there is no further improvement in neither player's best responses against provisional equilibrium mixed strategy. The following subsection describes our proposed SAC-AP approach for calculating best response oracles of the attacker and defender.
\subsection{SAC-AP}
In this paper, we propose soft actor-critic based deep RL for alert prioritization algorithm, i.e., SAC-AP, where both the attack and defense oracles use SAC-AP to compute pure strategy best response against opponent's mixed strategy. SAC-AP operates on continuous action spaces and computes an approximate pure strategy best response $\pi_v$ to an opponent using a mixed strategy $\sigma_{-v}$. For each player $v$, SAC-AP uses neural networks to represent policy network (actor), value network, and action value networks (critic). These networks are described below.
\subsubsection{Critic Network}
The two critic networks $Q_{1,2}(O_v,\alpha_v|\theta_v^{Q_{1,2}})$ map an observation and an action to a value with parameters $\theta_{v}^{Q_1}$ and $\theta_{v}^{Q_2}$, respectively. SAC uses two critics to mitigate positive bias in the policy improvement step that is known to degrade the performance of value-based methods. This technique also speeds up training and leads to better convergence. We train the two critic functions independently to optimize the critic’s objective function, $J_{Q_{i}}$, ($i = 1,2$) as:
\begin{equation}
    J_{Q_{i}} = E\big[\frac{1}{2}\big(Q^{'}-Q_i(\theta_{v}^{Q_i})\big)^2\big]
\end{equation}
where, $Q^{'} = R_v + \gamma E\big(\pi_v(\theta_{v}^{v^{'}})\big)$ and target network is used to estimate the state value instead of actual value network to improve stability. %$R_v$ represents player $v's$ reward and $\gamma$ is the discount factor.
To update the weights of the critic network with the gradient descent method, the loss function $J_{Q_{i}}$ is minimized as follows:
\begin{equation}
    \theta_{v}^{Q_i} \leftarrow \theta_{v}^{Q_i} - \lambda_Q\nabla J_{Q_i}
\end{equation}

\subsubsection{Value Network}
The state value function approximates the soft value. The value network $V(O_v|\theta_v^v)$ has parameters $\theta_v^v$, and maps the state to a value. The soft value, $J_{v}$ function is trained to minimize the squared residual error:
\begin{equation}
    J_v = E\big[\frac{1}{2}*\big( V^{'}-V(\theta_v^v)\big)^2\big]
\end{equation}
where, 
\begin{equation}
V^{'} =  E \big[Q-log\big( \pi_v(\theta_v^\pi)\big)\big] 
\end{equation}
and, 
\begin{equation}
Q =  \min \big[E\big(Q_1(O_v^k,\pi_v(\theta_v^\pi)\big), E \big(Q_2(O_v^k,\pi_v(\theta_v^\pi)\big)\big]
\end{equation}
Further, we use gradient descent \cite{alpaydin2020introduction} to train the weights of the value network and exponential moving average to update the target value network:
\begin{align*}
\theta_v^v \leftarrow \theta_v^v-\lambda_v\nabla J_v\\
\theta_v^{v^{'}} \leftarrow \tau\theta_v^v + (1-\tau)\theta_v^{v^{'}}
\end{align*}

\subsubsection{Action Network}
The actor-network $\pi_v(O_v|\theta_v^\pi)$ maps observation $O_v$ into an action with parameters $\theta_v^\pi$. It uses policy improvement to learn pure best strategy response $(\pi_v)$. The objective function, $J_\pi$, is defined as:
\begin{equation}
  J_\pi = E\big[-Q + \beta log\big(\pi_v(\theta_v^\pi)\big)\big]
\end{equation}
where, $\beta$ is the entropy regularization factor which controls exploration-exploitation, and 
\begin{equation}
Q = min\big[E\big(Q_1(O_v^k,\pi_v(\theta_v^\pi)\big), E \big(Q_2(O_v^k,\pi_v(\theta_v^\pi)\big)\big]
\end{equation}

Similar to the critic and value networks, We use gradient descent to minimize the above loss:
\begin{equation}
    \theta_v^\pi \leftarrow \theta_v^\pi-\lambda_\pi\nabla J_\pi
\end{equation}

Initially, all four neural networks are randomly initialized. Then, they are trained over multiple episodes. At the beginning of each episode, the opponent samples a deterministic policy $\pi_{-v}$ with its mixed strategy $\sigma_{-v}$. The networks are trained as follows. First, the actor generates an action using the $\epsilon$-greedy method. Player $v$ then executes the action $\alpha_v$ using policy $\pi_v$ and player $-v$ executes an action $\alpha_{-v}$ using policy $\pi_{-v}$. The player $v$ receives a reward based on the state transition. It stores the $k^{th}$ transition $<O_v^k, \alpha_v^k, r_v^k, O_v^{k+1}>$ into a memory buffer. Player $v$ then samples a minibatch, a subset of transitions randomly sampled from the memory buffer, to update the networks. After a fixed number of episodes, the resulting policy network $\pi_v(O_v|\theta_v^\pi)$ is returned as the parameterized optimal response to an opponent with mixed strategy $\sigma_{-v}$. The complete SAC-AP algorithm is described in Algorithm $1$ and all the relevant hyperparameters are listed in Table \ref{table3}.

\begin{table}[hbtp]\centering
 \caption{Hyperparameter configurations for SAC-AP experiments }
 \label{table3}
 \begin{tabular}{ |p{2 cm}|p{4 cm}|p{1 cm}| }
 \hline
 Hyperparameter & Description &Value \\
 \hline
 $\lambda_Q$ & Learning rate of critic networks & 0.002\\
\hline
$\lambda_v$ &Learning rate of value network &0.002\\
\hline
$\lambda_\pi$&Learning rate of actor network &0.001\\
\hline
$\gamma$ &Discount factor &0.95\\
\hline
$\beta$&Entropy regularization factor &0.5\\
\hline
$\tau$&Smoothing constant &0.01\\
\hline
$\epsilon_{max}$&Maximum epsilon value &1\\
\hline
$\epsilon_{discount}$&Epsilon discount factor &0.99\\
\hline
\end{tabular}
\end{table}
\begin{algorithm}
\caption{SAC-AP Algorithm: Compute the pure-strategy best response of player $v$ when its opponent takes mixed strategy $\sigma_{-v}$.}\label{alg}
\hspace*{\algorithmicindent} \textbf{Input}: The set of opponent’s pure strategies, $\Pi_{-v}$ and mixed strategy of the opponent, $\sigma_{-v}$;\\
 \hspace*{\algorithmicindent} \textbf{Output}: The policy network of player $v$, $\pi_v(O_v|\theta_v^\pi)$, the value network of player $v$, $V(O_v|\theta_v^v)$ and the critic networks of player $v$, $Q_{1,2}(O_v,\alpha_v|\theta_v^{Q_{1,2}})$ ;

\begin{algorithmic}[1]
\raggedright
\State Randomly initialize $\pi_v(O_v|\theta_v^\pi)$, $V(O_v|\theta_v^v)$ and $Q_{1,2}(O_v,\alpha_v|\theta_v^{Q_{1,2}})$;
\State Initialize replay memory $D$;
\For {$episode = 0,M-1$}
 \State Initialize the system state $<N^{(0)}, M^{(0)}, S^{(0)}>$;
 \State Sample opponent’s policy $\pi_{-v}$, with its mixed \hspace*{5mm}strategy $\sigma_{-v}$ over $\Pi_{-v}$;
  \For{k=0, k-1}
   \State With probability $\epsilon$ select random action $\alpha_v^{(k)}$;
   \State Otherwise select  $\alpha_v^{(k)}$ $=$ $\pi_v(O_v|\theta_v^\pi)$;
   \State Execute $\alpha_v^{(k)}$ and $\alpha_{-v}^{(k)} = \pi_{-v}(O_{-v}^{(k)})$, observe \hspace*{1cm} reward $r_v^k$ and transit the system state to $s^{k+1}$;
   \State Store transition $<O_v^k, \alpha_v^k, r_v^k, O_v^{k+1}>$ in $D$;
   \State Sample a random minibatch of $N$ transitions \hspace*{1cm}$<O_v^k, \alpha_v^k, r_v^k, O_v^{k+1}>$ from $D$;
   \State Set $Q=min\big[E\big(Q_1(O_v^k,\pi_v(\theta_v^\pi)\big)$, \hspace*{1cm}$E \big(Q_2(O_v^k,\pi_v(\theta_v^\pi)\big)\big]$;
   \State Set $J_\pi = E\big[-Q + \beta log\big(\pi_v(O_v|\theta_v^\pi)\big)\big]$;
   \State Set $V^
   {'} =  E \big[Q-log\big( \pi_v(\theta_v^\pi)\big)\big]$;
   \State Set $J_v = E\big[\frac{1}{2}*\big( V^{'}-V(\theta_v^v)\big)^2\big]$;
   \State Set $Q^{'} = r_v^k + \gamma*\pi_v(\theta_v^{v^{'}})$;
   \State Set $J_{Q_{1}} = \frac{1}{2}*E\big[\big( Q^{'}-Q_1(\theta_v^{Q_1})\big)\big]^2$;
   \State Set $J_{Q_2} =\frac{1}{2}*E\big[\big( Q^{'}-Q_2(\theta_v^{Q_2})\big)\big]^2$;
   \State $\theta_v^{Q_1}\leftarrow\theta_v^{Q_1}-\nabla J_{Q_1}$;
    \State $\theta_v^{Q_2}\leftarrow\theta_v^{Q_1}-\nabla J_{Q_2}$;
         \State $\theta_v^{\pi}\leftarrow\theta_v^{\pi}-\nabla J_{\pi}$;
            \State $\theta_v^{v}\leftarrow\theta_v^{v}-\nabla J_{v}$;
               \State $\theta_v^{v^{'}}\leftarrow \tau\theta_v^{v}+(1-\tau)\theta_ v^{v^{'}}$;
  \EndFor
 \EndFor
\State \textbf{return} player $v$'s policy network $\pi_v(O_v|\theta_v^\pi)$;
\end{algorithmic}
\end{algorithm}

\section{Results}
In this section, we present the experimental setup and the results. 

\subsection{Simulation setup}
The SAC-AP algorithm is implemented in Tensorflow \cite{abadi2016tensorflow}, an open-source library for training neural networks. We used Adam optimizer for learning the parameters of the networks. The values of various hyperparameters are shown in Table \ref{table3}. We investigate two case-studies for the experiments: (i) fraud detection \cite{link} and (ii) network intrusion \cite{link2}. We compare SAC-AP with several state-of-the-art and traditional alert prioritization methods including Uniform, GAIN \cite{laszka2017game}, RIO \cite{yan2018get}, Suricata \cite{link2} and recently proposed DDPG-MIX algorithm, which is based on the deep reinforcement learning. Furthermore, we consider both the cases when defender is aware and unaware of adversary's capabilities. The expected loss of the defender i.e., the negative utility of the defender (equation \ref{eqq}) is used as the performance metric to evaluate the performance of the proposed approach and its comparison with other state-of-the-art alert prioritization methods.

\subsection{Case study 1 : Fraud detection}
\begin{figure}[h]
   \begin{center}
    \includegraphics[scale=0.37]{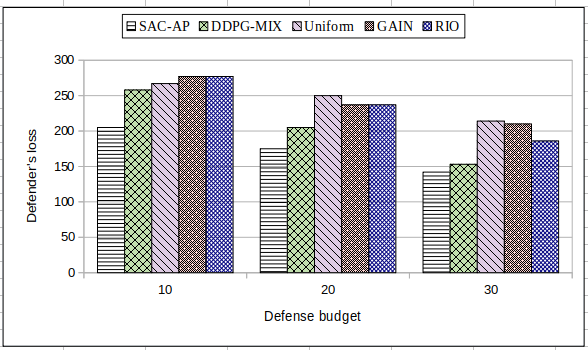}
    \end{center}
    \caption{Usecase: Fraud Detection - Performance comparison of alert prioritization techniques for different budget values when adversary's attack budget is fixed to 2.}
    \label{f_1}
\end{figure}

\begin{figure}[h]
   \begin{center}
    \includegraphics[scale=0.37]{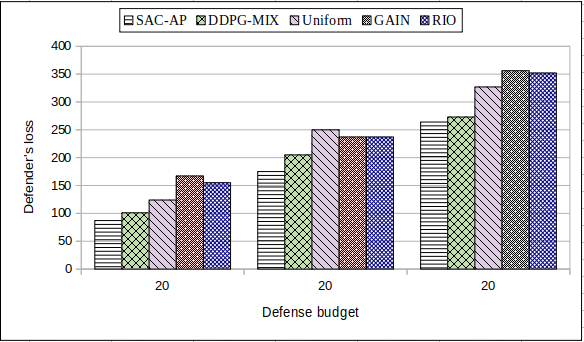}
    \end{center}
    \caption{Usecase: Fraud Detection - Performance comparison of alert prioritization techniques for different adversary's budget values when defender's defense budget is fixed to 20.}
    \label{f_2}
\end{figure}

Our first case study includes learning based fraud detector along with fraud dataset \cite{link}. There are 284,807 credit card transactions in the fraud dataset, with 482 of them being fraudulent. A vector of 30 numerical attributes represents each transaction and each feature vector has a binary label that indicates the transaction type. First we consider the two cases when the defender is aware of the adversary's attack budget. Figure \ref{f_1} presents the performance comparison of different alert prioritization techniques for different budget values when the adversary's attack budget is fixed to 2. Our results show that SAC-AP performs 20.54\%, 14.63\%, and 7.18\% better than DDPG-MIX for budget values 10, 20 and 30 respectively. The gains are even higher when SAC-AP is compared with Uniform, RIO and GAIN. For fraud detection, Suricata is not used since it is specifically designed for intrusion detection and therefore it is used in the next subsection for the network intrusion data set. Further, Figure \ref{f_2} presents performance evaluation for different adversary's attack budgets when the defense budget is fixed to 20. For this case, our results show that SAC-AP performs 13.86\%, 14.63\%, and 3.29\% better than DDPG-MIX for attack budget 1, 2 and 3 respectively.

\begin{figure}[h]
   \begin{center}
    \includegraphics[scale=0.37]{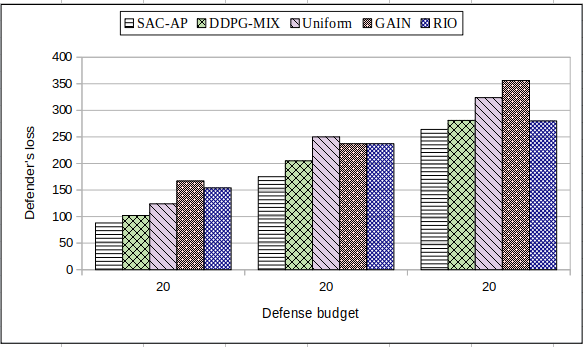}
    \end{center}
    \caption{Usecase: Fraud Detection - Performance comparison of alert prioritization techniques for different adversary's attack budget values when defender's defense budget is fixed to 20 and estimated attack budget is 2.}
    \label{f_3}
\end{figure}

Second, we consider the case when the defender is not aware of the adversary's attack budget. Figure \ref{f_3} presents the performance comparison of different alert prioritization techniques when the defense budget is fixed to 20. Since, the defender is unaware of the adversary's attack budget, we assume that it estimates it to 2. Figure \ref{f_3} shows results when the actual attack budgets are 1, 2 and 3. For this case, our results show that SAC-AP performs 13.72\%, 14.63\%, and 6.04\% better than DDPG-MIX for the actual attack budget values of 1, 2 and 3 respectively.
%\begin{figure}[hbt!]
%   \begin{center}
%    \includegraphics[scale=0.60]{TB2.png}
%    \end{center}
%    \caption{Usecase: Fraud Detection - Performance comparison of alert prioritization techniques for different adversary's policies when adversary's attack budget is fixed to 2 and defender's defense budget is fixed to 20.}
%    \label{picture5}
%\end{figure}

%\begin{figure}[hbt!]
%   \begin{center}
%    \includegraphics[scale=0.60]{TB1.png}
%    \end{center}
%    \caption{Usecase: Network Intrusion Detection - Performance comparison of alert prioritization techniques for different adversary's policies when adversary's attack budget is fixed to 120 and defender's defense budget is fixed to 1000.}
%    \label{picture9}
%\end{figure}

\subsection{Case study 2 : Network intrusion detection}
\begin{figure}[hbt!]
   \begin{center}
    \includegraphics[scale=0.36]{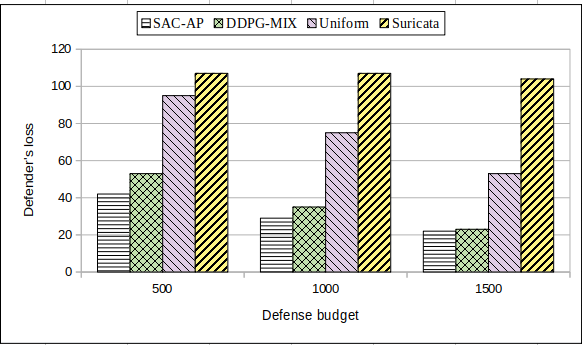}
    \end{center}
    \caption{Usecase: Network intrusion detection - Performance comparison of alert prioritization techniques for different budget values when adversary's attack budget is fixed to 120.}
    \label{i_1}
\end{figure}

\begin{figure}
   \begin{center}
    \includegraphics[scale=0.37]{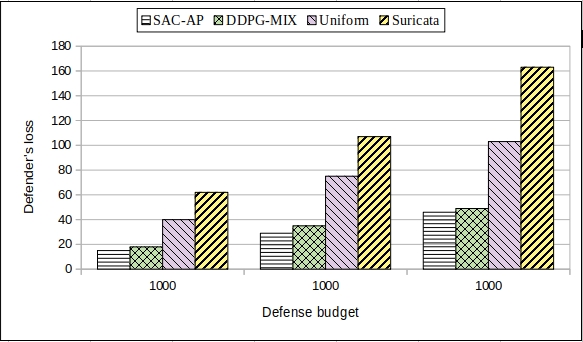}
    \end{center}
    \caption{Usecase: Network Intrusion Detection - Performance comparison of alert prioritization techniques for different adversary's budget values when defender's defense budget is fixed to 1000.}
    \label{i_2}
\end{figure}

Our second case study includes suricata IDS \cite{link2} an open source NIDS along with CICIDS2017 dataset \cite{sharafaldin2018toward}. This dataset was developed in 2017 by Canadian institute for cyber security (CIC). It is a large dataset that contains around 3 million network flows in different files. It contains 2,830,743 records where each record contains 78 different features. First, we consider two cases when the defender is aware of the adversary's attack budget. Figure \ref{i_1} presents the performance comparison of different alert prioritization techniques when the adversary's attack budget is fixed to 120. Our results show that SAC-AP performs 20.75\%, 17.14\%, and 4.54\% better than DDPG-MIX for the defense budget values of 500, 1000 and 1500, respectively. The gain is even higher when compared with other IDS including Uniform and Suricata. For this case, RIO and GAIN are not used since they are not suitable for the large size of CICIDS2017 dataset and are computationally expensive. Figure \ref{i_2} presents the gain of using SAC-AP for fixed defense budget of 1000 and for different adversary's attack budget values. 

\begin{figure} 
   \begin{center}
    \includegraphics[scale=0.37]{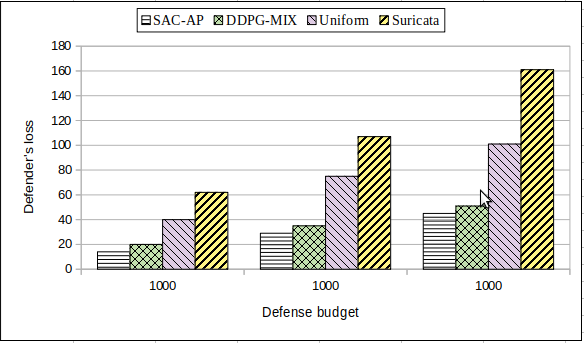}
    \end{center}
    \caption{Usecase: Network Intrusion Detection - Performance comparison of alert prioritization techniques for different adversary's attack budget values when defender's defense budget is fixed to 1000 and estimated attack budget is 120.}
    \label{i_3}
\end{figure}

We have also considered the case when the defender is unaware of the adversary's attack budget. Figure \ref{i_3} presents the scenario when the defense budget is fixed to 1000 and the defender is unaware of the attacker's budget. We assume that it estimates it to 120 and evaluate the defender's loss for the attack budget of 60, 120 and 180.  Specifically, when the actual attack budget value is 60, the defender is over estimating while when the actual attack budget value is 180, the defender is underestimating. However, for all these cases, SAC-AP performs 30\%, 17.14\%, and 11.76\% better than DDPG-MIX.

%\subsection{Different attacker policies for two use-cases}

%Finally, Figure \ref {picture5} and Figure \ref{picture9}  presents the performance comparison of SAC-AP with other alert prioritization methods when the attacker utilizes different policies: Uniform, Greedy, DDPG-M

%Uniform, and Greedy instead of DDPG-MIX and SAC-AP on which the defender is trained. We used the DDPG-MIX attacker to be consistent. We can see that our method outperforms all the other algorithms when the attacker uses SAC-AP and DDPG-MIX policy. Our approach even outperforms a defender trained by the DDPG-MIX algorithm against the same attacker. This shows the robustness of our model against an intelligent attack policy. Our approach also outperforms all the baselines when the attacker is using Uniform policy but performs worse compared to other baselines when the attacker switches to Greedy policy. However, a rational adversary could cause Uniform, RIO, and GAIN (for fraud detection) and Uniform and Suricata (for intrusion detection) to degrade by nearly 26\%, whereas SAC-AP is quite robust to such adversaries. 

Our results show that the proposed approach outperforms several state-of-the-art and traditional alert prioritization techniques for the above mentioned case studies for both the scenarios when the defender is aware and unaware of adversary's attack budget.

\section{Conclusion}
In this work, a novel alert prioritization method SAC-AP, based on deep reinforcement learning and double oracle framework is proposed. We implemented SAC-AP, which is based on maximum entropy reinforcement learning framework. Our proposed approach outperforms various state-of-the art alert prioritization methods and achieve up to 30\% gain when compared to recently proposed RL based DDPG-MIX algorithm. We conduct extensive experiments for two use cases: fraud and intrusion detection, and present the efficacy of the proposed approach in obtaining robust alert investigation policies. Future work includes the comprehensive study of investigating the performance of SAC-AP on the scenarios with multiple attackers.

\bibliographystyle{ieeetr}
\bibliography{reference.bib}
\end{document}